\documentstyle[twoside,fleqn,espcrc2,epsfig]{article}

\newcommand{\micr}{\; \mu {\mathrm m}}
\newcommand{\cm}{\; {\mathrm cm}}

\title{Monitoring the Stability of the ALEPH Vertex Detector}

\author{
  G.Sguazzoni$^{\mathrm i}$, 
  D.Creanza, M.de~Palma, G.Maggi, G.Raso, G.Selvaggi, L.Silvestris,
  P.Tempesta\address{Dip. di Fisica, 
    INFN Sezione di Bari, 70126 Bari, Italy},
  M.Burns, M.Frank, P.D.Maley, M.Morel, A.Wagner\thanks{Now at
Schweizerischer Bankverein, Basel, Switzerland.}\address{European
  Laboratory for Particle Physics (CERN), CH 1211  
    Geneva 23, Switzerland},
  E.Focardi, G.Parrini, E.Scarlini\address{Dip. di Fisica
  dell'Universit\`a e INFN Sezione 
    di Firenze, 50125 Firenze, Italy},
  A.Halley, V.O'Shea, C.Raine\thanks{Deceased.}\address{Dept. of
  Physics and Astronomy, University of Glasgow, 
    Glasgow G12 8QQ, United Kingdom},
  G.Barber, W.Cameron, P.Dornan, D.Gentry, A.Moutoussi, 
  J.Nash, D.Price, A.Stacey, L.W.Toudup\address{Dept. of Physics,
  Imperial College, London SW7 2BZ, United 
    Kingdom}, 
  M.I.Williams\address{Dept. of Physics, University of Lancaster,
    Lancaster LA1 4YB, United Kingdom},
  M.Billault, P.E.Blanc, A.Bonissent, G.Bujosa, D.Calvet, J.Carr,
  P.Coyle, C.Curtil, J.J.Destelle, C.Diaconu, D.Fouchez, P.Karst, P.Payre,
  D.Rousseau, M.Thulasidas\address{CPPM, Facult\'e des
    Sciences de Luminy, IN2P3-CNRS, 13288 Marseille, France},
  H.Dietl, G.Ganis, H.G.Moser, R.Settles, H.Seywerd,
  G.Waltermann\address{Max-Planck-Institut f\"ur Physik,
  Werner-Heisenberg-Institut, 
    80805 M\"unchen, Germany}, 
  F.Bosi, C.Bozzi, R.Dell'Orso, A.Profeti, G.Rizzo,
  P.G.Verdini\address{Dip. di Fisica dell'Universit\`a, INFN Sezione
  di Pisa e Scuola Normale Superiore, 56010 Pisa, Italy},
  J.P.Bizzell, J.C.Thompson\address{Particle Physics Dept., Rutherford
  Appleton Laboratory, 
    Chilton, Didcot, Oxon OX11 OQX, UK},
  S.Black, J.Dann, H.Y.Kim, N.Konstantinidis,
  G.Taylor\address{Institute for Particle Physics, University of 
    California at Santa Cruz, Santa Cruz, CA 95064, U.S.A.},
  L.Bosisio\address{Dip. di Fisica dell'Universit\`a e INFN Sezione di Trieste,
  34127 Trieste, Italy}, 
  J.Rothberg, S.Wasserbaech\address{Experimental Elementary Particle
  Physics, University of 
    Washington, WA 98195 Seattle, U.S.A.},
  S.Armstrong, P.Elmer, J.Walsh\address{Dept. of Physics, University
  of Wisconsin, Madison, WI 53706, 
    U.S.A.} 
}  
\begin{document}
        
\begin{abstract}
  The ALEPH Silicon Vertex Detector features an optical fibre
  laser system to monitor its mechanical stability. 
  The operating principle and the general
  performance of the laser system are described. 
  The experience obtained during 1997 and 1998 operations confirms the
  important role that such a system can have with respect to the
  detector alignment requirements. In particular, the laser system  
  has been used to monitor short-term temperature-related effects and 
  long-term movements. These results and a description of
  the laser-based alignment correction applied to the 1998 data are presented.
\end{abstract}
\maketitle
\section{Introduction}

The ALEPH Silicon Vertex Detector (VDET) has an active length of 40 cm
and consists of two concentric cylindrical layers of 144 
micro-strip silicon detectors of $\sim 5.3 \times 6.5 \cm^2$ with
double-sided readout. Six of them are glued together and
instrumented with readout electronics on each end to form the VDET
elementary unit ({\em face}). 
The inner layer ($\sim 6.3 \cm$ radius) is formed by 9 faces, the
outer layer ($\sim 10.5 \cm$ radius) consists of 15  
faces. Strips on $p^+$-side run parallel to $z$ axis in the ALEPH
reference system, allowing $r \phi$ coordinates measurement; $n$-side
strips, running normal to $z$ axis, allow $z$ coordinate measurement.
A detailed description of VDET can be found elsewhere~\cite{ettore}.

\section{Laser system}

VDET features a laser system to monitor mechanical
stability with respect to the external tracking devices.
A large movement ($> 20 \micr$) of the VDET during data-taking 
could degrade significantly the $b$-tagging performance. 
It is thus very important to keep this aspect under strict control. 
This is especially true at LEP2, where the detector alignment is 
performed on an initial sample of Z events and monitoring using the 
high energy data itself is difficult due to the reduced event rates. 

\begin{figure}[t]
    \leavevmode
    \epsfig{file=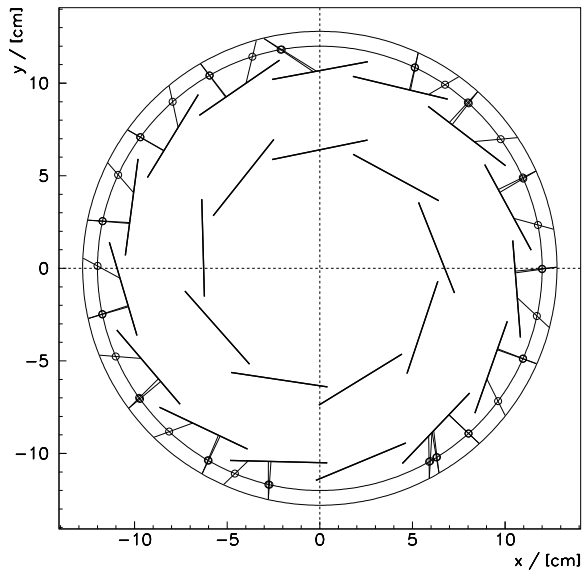, height=7.5cm}
    \caption{VDET $r \phi$ section with laser beams.}
    \label{fig:vdetsec}
\end{figure}

\begin{figure}[t]
    \leavevmode
    \epsfig{file=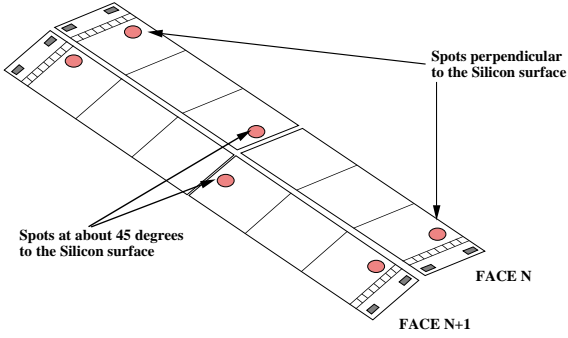, width=7.5cm}
    \caption{VDET faces with laser spots.}
    \label{fig:vdetface}
\end{figure}

The system uses infrared light ($\lambda = 904\; {\mathrm nm}$) from two
pulsed laser diodes with an output power of $6\; {\mathrm W}$ each 
and a pulse length of $50\; {\mathrm ns}$.
The light is distributed via optical fibres to prisms and lenses 
attached to the inner wall of the Inner Tracking Chamber (ITC, the closest
outer detector). The lenses focus 44 light spots 
on 14 of the 15 VDET outer faces\footnote{One face is not equipped because
  the mechanical structure nearby does not allow the optic
  installation.}, normally 3 spots per face, two spots
close to the ends of the face and one spot about in the middle of the face.
All laser beams are nominally parallel to the $xy$ plane; laser
beams at the VDET ends are normal to the silicon surface while laser beams
that point to the central wafers have an azimuthal incident angle close
to 45 degrees, in order to be sensitive to movements normal to the face plane. 
Information on the VDET displacements
with respect to the ITC are obtained by
monitoring the laser beam impact position on the silicon wafers ({\em laser
spot}) versus time: 
the $r \phi $ spot position is sensitive only to displacements in the $xy$
plane and the $z$ spot position is sensitive only to displacements along 
the $z$ direction. 
An $xy$ section of the VDET indicating the nominal laser beam positions
is shown in Fig.~\ref{fig:vdetsec}. In Fig.~\ref{fig:vdetface} a 
sketch of the typical laser spot positions on a face is shown.
For a detailed description of the VDET laser system see~\cite{bib2}.

The laser system operates during standard data taking: the laser trigger, 
synchronized with the beam-crossing signal, is fired
once per minute (approximately once per $\sim 100$ physics events) and
the laser event is acquired as a standard physics event.
After the installation 5 spots out of 44 spots were absent,
probably due to misalignment of the optics or breakage of the optical fibre.
For the remaining spots, the efficiency is essentially 100\%
due to the large pulse height of a spot cluster. 
During 1997 (1998) $\sim 62000$ ($\sim 129000$) laser events were
collected. 

\section{Analysis of displacements}

Each laser spot position is reconstructed event-by-event using 
a standard centre-of-gravity algorithm.
A deviation, $\Delta$, is defined for each event and for each spot as
the difference between the actual spot position and a nominal position.

\begin{figure}[h]
  \begin{center}
    \leavevmode
    \epsfig{file=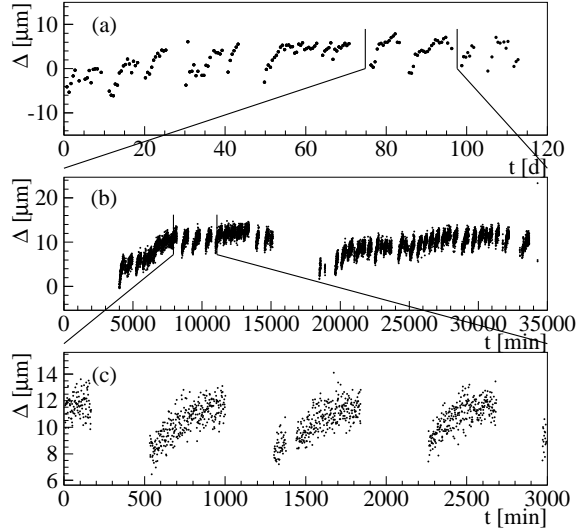, width=7.5cm}
    \caption{Typical spot time-chart with different horizontal 
     time scale to display the (a) long term, (b) medium term, (c) 
     short term.}
    \label{fig:scantau}
  \end{center}
\end{figure}

The raw deviation for a typical spot as a function of time shows various features: 
a long term effect (Fig.~\ref{fig:scantau}~(a)), 
a medium term effect, present after quite long shutdown periods (Fig.~\ref{fig:scantau}~(b)),  
and a short term effect with a timescale of a few hours (Fig.~\ref{fig:scantau}~(c)).

\subsection{Short and Medium Term Effects} 

\begin{figure}[t]
    \leavevmode
    \epsfig{file=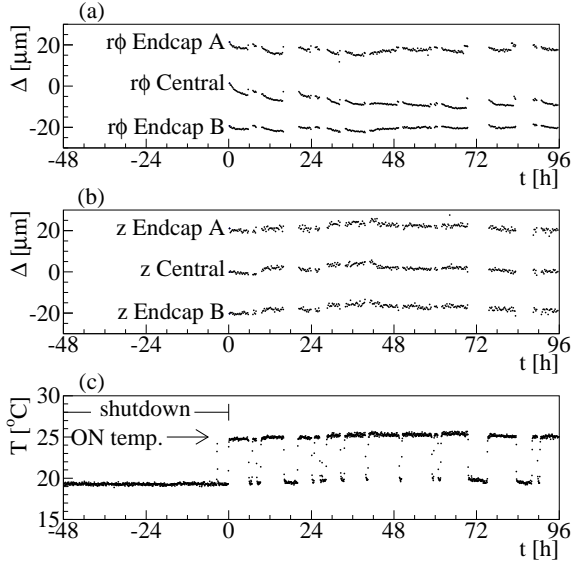, width=7.5cm}
    \caption{Spot deviations of a face: (a) $r\phi$ side;
    (b) $z$ side; (c) VDET temperature.  Side A (side B) 
    deviations have been shifted by 20$\micr$ ($-20 \micr$) to fit the
    same plot.} 
    \label{fig:short}
\end{figure}

The $r\phi$ and $z$ deviations of the three spots of a face 
are plotted in Fig.~\ref{fig:short}~(a) and (b) 
for a four day period after a long shutdown. Also  
shown is a time chart of the VDET temperature over the same 
period (Fig.~\ref{fig:short}~(c)). 
The $r\phi$ side shows clear short and medium term effects 
correlated with the temperature variations.
In order to reduce the probability of radiation damage,  
VDET is completely ON only when LEP is in ``stable beams''.
The $r\phi$ central spot, which is at $45^\circ$ and therefore
sensitive to radial motion, 
shows the biggest deviation and is consistent with the expected face 
bowing due to the bimetallic effect (the face has a kevlar and carbon
fibre beam 
glued on the silicon to ensure mechanical rigidity).
%
The medium term effect is visible in~Fig.~\ref{fig:short}~(a) for the 
deviation of the central spot. It seems that ``equilibrium'' is only 
reached after about two days after a long shutdown. 
This medium term effect is also thought to be due to the face bowing. 
During standard running, OFF periods are short ($\sim 1 \div 2$ hours) 
and recovery from thermal expansion is partial; during long
shutdowns (more than $\sim 1$ day) the recovery is complete and after
that a certain time is needed to reach again the normal warming-cooling cycle. 

The maximum bowing sagitta $S$ and the time constant $\tau$
involved in these temperature related phenomena can be estimated by
fitting the following parametrisation over the 45$^\circ$ spot
deviation $\Delta$ versus time $t$: 
\begin{equation}
\Delta(t) = S \cdot \left( 1 - e^{-(t-t_0)/\tau} \right).
\label{eq:param}
\end{equation}
The fits have been done over a typical ON
period just after a long shutdown (with $t_0 = 0$) and over a
standard ON period far from a long shutdown (with free $t_0$).
Depending on the face, the sagitta ranges from $\sim 5$ to $12 \micr$ 
with a mean of $8\micr$ and the time constant $\tau$ varies
from $\sim 130$ to 280~min with a mean of 160~min.
For the tracking performance the short term displacements are small 
enough to be neglected, especially as they mainly affect the radial direction.

For the short term case the width of the residual distribution of 
data points with respect to the parametrisation in Eq.~\ref{eq:param}
is an estimate of the spatial resolution for a single event.
Thanks to the large pulse height and the fact the cluster extends over 
more than one readout strip, the resolution is typically $0.5$ to 
$1.5 \micr$. 

\subsection{Long-term Effects}

\begin{figure}[t]
  \begin{center}
    \epsfig{file=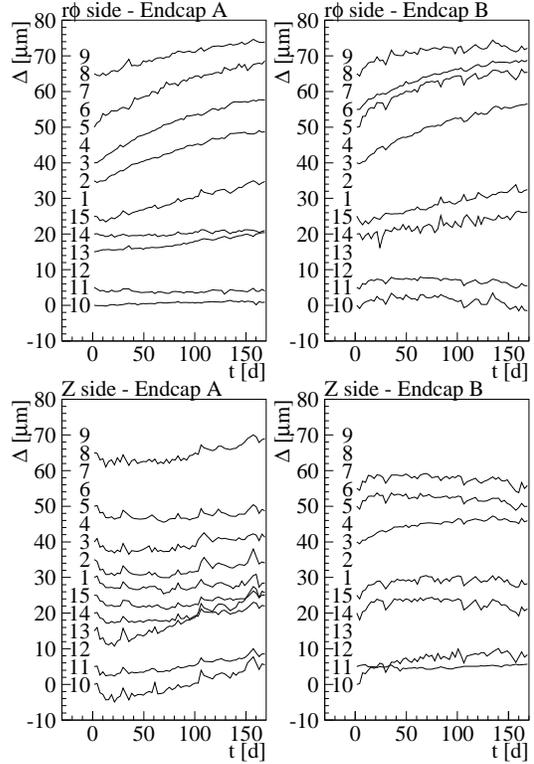, width=6.9cm}
    \caption{1998 long term $r\phi$ and $z$ deviations time chart for
    the good endcap spots. The number displayed is the face
    number. The initial values of the deviations are shifted to allow 
    display on the same plot.} 
    \label{fig:rawtc}
  \end{center}
\end{figure}

\begin{figure}[t]
  \begin{center}
    \leavevmode
    \epsfig{file=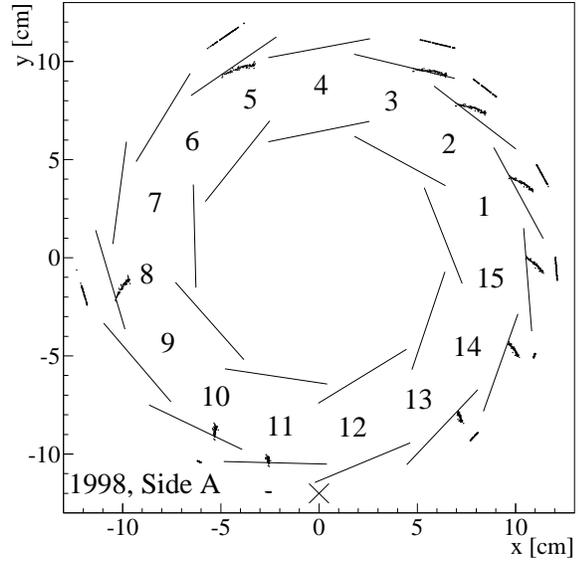, width=7.5cm}
    \caption{ Fitted spot displacements for the endcap A spots: the
    points represent the spot position during time. 
    The displacement is amplified by a factor of 1000. 
    The small segments, plotted outside the outer layer, are
    the actual raw deviations of the corresponding spot (with the same
    magnification). The cross is the probable rotation centre.} 
    \label{fig:reconstruction}
  \end{center}
\end{figure}

The long term effect is visible in Fig.~\ref{fig:rawtc}, where the
spot deviations with respect to initial values are plotted versus
time for the entire 1998 data taking period. Some spots are not displayed
because they are missing or are not used in the analysis due to
inadequate pulse height or unusual cluster profile. 
The $r\phi$ side deviations show a systematic long term drift that depends on
face number (see Fig.~\ref{fig:reconstruction} for face numbering convention). 
The size of the $xy$ displacements are as large as $\sim 20 \micr$
and are larger than the single hit resolution on a charged track ($\sim 10 \micr$). 
On the contrary, the $z$ side deviations are smaller and quite similar. 

During 1997 and 1998 this unexpected behaviour has been studied,
parametrised and an alignment correction has been implemented.

\begin{figure}[t]
  \begin{center}
    \leavevmode
    \epsfig{file=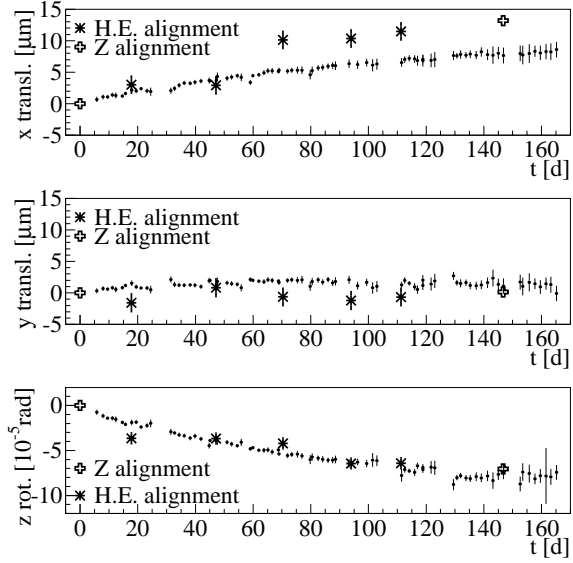, width=7.5cm}
    \caption{Long term $xy$ alignment parameters relative to 
    the initial ${\mathrm Z}$ calibration run. The same parameters
    obtained from the alignments using charged tracks are also plotted.} 
    \label{fig:parameters}
  \end{center}
\end{figure}
The laser spots do not provide enough information to fully reconstruct the
displacements of each individual face, nevertheless the observed deviations do
not seem to be due to an independent motion of the faces. The
observed long term deviations are thus parametrised 
assuming that VDET is moving as a rigid body with respect
to the initial position, given by the alignment procedure with tracks performed  
at the beginning of data taking in a run at the ${\mathrm Z}$ resonance.   
The deviation of a spot is expressed as a function of
the standard parameters used to describe the motion of rigid bodies,
consisting of 3 angles and 3 translation vectors. These are then
extracted by a fit procedure that minimises the squared differences between 
the observed and the predicted deviation of the spots.

The fit has been performed, with 3 out of the 6 parameters fixed to zero: 
the two rotation angles about the $x$ and $y$ axes, having found
negligible displacement in $xz$ and $yz$ planes, and the translation 
along $z$ because the observed $z$ displacement is more consistent
with a deformation rather than a global displacement. 
The result is shown in Fig.~\ref{fig:parameters}, where the three free 
parameters ($x$ translation, $y$ translation and $z$ rotation) are plotted 
versus time starting from the initial ${\mathrm Z}$
run. A pictorial view of the corresponding VDET displacement
is shown in Fig.~\ref{fig:reconstruction}; it is consistent with a rotation 
around the bottom face.  

\begin{figure}[t]
  \begin{center}
    \leavevmode
    \epsfig{file=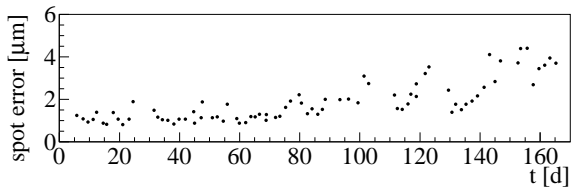, width=7.5cm}
    \caption{The average single spot standard deviation from the fit after 
    imposing the chi-square per degree of freedom to be 1.} 
    \label{fig:errors}
  \end{center}
\end{figure}

For the fit, the deviations have been averaged over a period of two days. 
The errors on the parameters are estimated ``a posteriori'' by forcing
the $\chi^2_{min}$ to be equal to the number of degrees of freedom; the
resulting uncertainty on the single spot deviation is plotted in
Fig.~\ref{fig:errors} versus time.  
Although the single spot spatial resolution is better than that given in 
Fig.~\ref{fig:errors}, the rigid body assumption does not take into account possible 
structural deformations, misalignment of the fibres or independent motion of 
the faces.  
    
The time dependent motion of the VDET extracted from the laser system 
has also been independently confirmed using charged tracks in 
data acquired at high energy and also during a run at
the ${\mathrm Z}$ resonance taken at the end of the 1998 data taking. 
To cope with the low statistics during the high energy running, 
the alignments were performed  averaging over a $\sim 1$ month period. 
The values for the alignment parameters so obtained are also plotted 
in Fig.~\ref{fig:parameters} and are generally in good agreement. 
In the $x$ translation, the alignment points are shifted with respect
to laser points from about day 60.
This is correlated with a beam loss into the Time Projection Chamber
field cage that caused a measurable (10$\micr$) relative
displacement of the TPC with respect to the ITC. This emphasises
that the alignment obtained directly from the data is also sensitive
to any time dependence of the alignment of the other tracking
chambers.

\begin{figure}[t]
  \begin{center}
    \leavevmode
    \epsfig{file=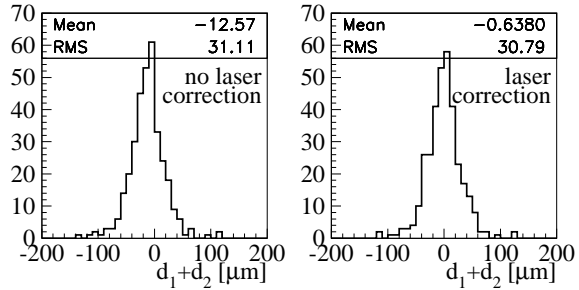, width=7.5cm}
    \caption{Distribution of sum of impact parameters of the two
    tracks in dimuon events for final calibration run before and after
    laser correction to the alignment.} 
    \label{fig:sumd0}
  \end{center}
\end{figure}

Based on the alignment parameters extracted from the laser system, 
a time dependent correction to the initial ${\mathrm Z}$ alignment has 
been applied to all the 1998 data. Fig.~\ref{fig:sumd0} shows the 
distribution of the sum of the impact parameters of the two muons in 
dimuon events before and after applying the alignment correction. 
Without the correction the mean of the 
distribution is shifted by $13 \micr$, when the correction is applied 
the distribution is centered close to zero as expected. 

The observed rotation of the VDET may be explained as follows:
the VDET is supported by two ``feet'' which slot into two long metal 
rails located at the top and bottom of the ITC cylinder. 
It seems that the VDET is rotating around the lower foot (small cross 
in Fig.~\ref{fig:reconstruction}). Although the lower foot is rigid 
(made of metal), the upper foot had to be ``springy'' (made of plastic) 
in order to allow for distance variations between the top and bottom rails as 
the VDET slides along the rails during the installation process. It is thought 
that the plastic foot suffers some small deformation. 
  
\section{Conclusions}

The laser based alignment system for the ALEPH vertex detector is operational.
It allows reliable and high precision monitoring of the VDET position during 
data taking. It has allowed the study of small temperature related motions and 
revealed a long term rotation of the VDET. An alignment correction based on the 
information from this system has been successfully applied.



\end{document}